# Crystalline-to-Crystalline Phase Transition between Germanium Selenide Polymorphs with High Resistance Contrast

*Joonho Kim[1], Kihyun Lee[1], Joong-Eon Jung[1], Han Joo Lee[1], Seongil Im[1], and Kwanpyo Kim[1,*]*

[1]Department of Physics, Yonsei University, Seoul 03722, Korea.

*Address correspondence to K.K. (kpkim@yonsei.ac.kr)

**Abstract**

**Understanding phase transitions between crystalline phases of a material is crucial for both fundamental research and potential applications such as phase-change memory. In this study, we investigate the phase transition between GeSe crystalline polymorphs induced by either global annealing at moderate temperatures or localized laser-induced heating. The highly conductive γ-GeSe transforms into semiconducting, single-crystalline α-GeSe while preserving a well-aligned crystal orientation. The distinct structural and electronic properties at the γ-GeSe/α-GeSe interface were investigated by transmission electron microscopy analysis. We propose that the clustering of Ge vacancies in the γ-GeSe phase at elevated temperatures is a key mechanism driving the transition, leading to the formation of α-GeSe through the segregation of a minor $GeSe_2$ phase. Furthermore, we observe a high electrical resistance contrast of approximately $10^7$ between γ-GeSe and α-GeSe, underscoring the potential of GeSe as a model polymorphic system for electronic applications, including phase-change memory.**

Phase transition is one of the fundamental phenomena in materials science, enabling researchers to tailor a material's structural, electronic, and magnetic properties.[1-7] In contrast to conventional three-dimensional materials, low-dimensional materials often exhibit distinctive and unique pathways for polymorphism and related phase transition. In particular, numerous layered two-dimensional crystals demonstrate polymorphism, supporting several stable phases with distinct chemical bonding and stacking configurations.[8-12] For instance, transition metal dichalcogenides such as $MoS_2$ and $MoTe_2$ exhibit multiple polymorphs - including semiconducting (2H) and metallic (1T and 1T′) phases - which can be exploited for developing new electronic devices via precise phase patterning.[13-18] Additionally, polymorphic transitions between different crystalline states have recently gained attention due to their potential to provide significant changes in material properties without involving amorphous intermediates.[19-22] Such transitions offer practical advantages by circumventing the structural instability and performance limitations commonly associated with amorphous phases.[23-27]

Group-IV monochalcogenides (MX, M = Ge, Sn; X = S, Se, Te) are promising materials for various applications including thermoelectricity and phase transition.[19,20,28-31] Among these, germanium selenide (GeSe) is a unique material due to the existence of stable polymorphs under normal temperature and pressure.[32] Among these, α-GeSe is the most prevalent polymorph, sharing a similar crystal structure with other sulfur- and selenium-based monochalcogenides. α-GeSe exhibits *p*-type semiconducting properties with a suitable bandgap ($E_g$ ~1.2 eV) for photoelectronic applications.[33,34] In contrast, γ-GeSe is an emerging polymorph of GeSe that shows significant differences from α-GeSe both in chemical bonding and electrical properties.[35-37] γ-GeSe is expected to be a narrow-gap semiconductor ($E_g$ ~0.4 eV), and it has been recently revealed that its unique intralayer bonding configuration induces a high level of Ge vacancies in the crystal.[36,38] These vacancies result in the high electrical conductivity of the material, a characteristic also found in tellurium-based phase-change materials. The substantial contrast in electrical resistance between these GeSe polymorphs suggests potential applications based on crystalline-to-crystalline phase transitions.

In this study, we demonstrated a crystalline-to-crystalline phase transition from γ-GeSe to α-GeSe with a significant contrast of electrical resistance and photoresponsivity. Encapsulated γ-GeSe was transformed into α-GeSe through global thermal annealing or laser irradiation. Transmission electron microscopy (TEM) measurements confirmed that the transformed α-GeSe is a single crystal with a well-aligned crystal orientation relative to the original γ-GeSe. We propose that the mechanism underlying this phase transition involves Ge vacancy clustering and the segregation of minor $GeSe_2$ phase within the crystal. Notably, the electrical resistance of the transformed α-GeSe is approximately $10^7$ times higher than the original γ-GeSe. Moreover, α-GeSe exhibited a robust photoresponse, which was absent in γ-GeSe. The significant contrast in electrical and optical properties, combined with the well-aligned crystal structure resulting from this phase transition, highlights GeSe polymorphs as promising candidates for future electronic devices and phase-change memory applications.

## Results and Discussion

**Figure 1a** illustrates the crystal structures of the representative polymorphs of GeSe, α-GeSe and γ-GeSe. α-GeSe features an orthorhombic puckered structure with covalent bonding, commonly observed in group-IV monochalcogenides composed of relatively light chalcogen elements. It has an anisotropic crystal structure with distinct zigzag (ZZ) and armchair (AC) directions. In contrast, γ-GeSe exhibits an isotropic hexagonal structure with a four-atom-thick layer, and a unique local bonding configuration violating the *8-N* rule and deviating from regular covalent bonding.[35,36] The observed bonding characteristic induces a high level of Ge vacancies and *p*-doping in γ-GeSe, resulting in significant differences in its electrical properties compared to α-GeSe. We synthesized γ-GeSe crystals by vapor–liquid–solid (VLS) growth on $SiO_2$/Si substrates coated with Au (see **Methods** for details). Compared to previously reported α-GeSe synthesis using the VLS method,[39] our approach involved the addition of iodine crystals to GeSe powder and the use of a slightly higher synthesis temperature, which enabled the preferred growth of γ-GeSe.

We investigated the phase transition between these polymorphs through thermal annealing. To prevent evaporation during thermal annealing,[15] γ-GeSe was encapsulated with graphite or h-BN, which are stable at high temperatures. Via global encapsulation annealing over approximately 530 °C, γ-GeSe was completely transformed into α-GeSe (**Figure 1b**). **Figure 1c** displays the Raman spectra before and after thermal annealing. Before annealing, five characteristic Raman peaks corresponding to γ-GeSe were clearly observed. After annealing, these peaks vanished completely and the Raman spectrum of α-GeSe was observed. We further examined the phase transition conditions by varying the annealing temperature and holding time at target temperature. The furnace temperature was ramped up to the target value in 30 minutes, maintained at target temperature, and then allowed to cool naturally to room temperature (**Figure S1a**). Each point in **Figure 1d** indicates the resulting polymorph of the encapsulated sample after thermal annealing. Additionally, we varied the holding time from 0 to 390 minutes and observed no significant dependence (**Figure S1b**). Our results confirm that the phase transition to α-GeSe occurs above approximately 530 °C, with slight variations in the transition temperature. The observed sample-to-sample variation in transition temperature is attributed to differences in encapsulation quality of the samples. Since γ-GeSe is not atomically thin, the encapsulation via graphite or h-BN is not always perfect, leading to different annealing environments and occasional partial material loss during thermal treatment.

To determine the crystallinity and orientation of the transformed α-GeSe crystal, we obtained selected area electron diffraction (SAED) patterns and measured angle-dependent optical contrast (**Figure S2**). The SAED pattern of the annealed sample matched the orthorhombic symmetry of α-GeSe, confirming its single-crystalline nature (**Figure S2a, b**). Additionally, we measured the optical contrast as a function of the angle between the ZZ direction of α-GeSe and the light polarization. The optical contrast exhibited a clear 180° periodicity, with maximum intensity observed when the ZZ direction was parallel to the light polarization (**Figure S2c, d**). Moreover, angle-dependent optical contrast plots obtained from different positions displayed similar angular dependencies, verifying that the crystal orientation of α-GeSe is consistent throughout the sample (**Figure S2e, f**).

We also investigated the relationship between the crystal orientations of the transformed α-GeSe and the original γ-GeSe. **Figure 1e** shows a TEM image of γ-GeSe prior to the phase transition, with the corresponding SAED pattern displaying hexagonal symmetry and three ZZ directions in the γ-GeSe crystal (**Figure 1f**). After the phase transition, the SAED pattern of single-crystalline α-GeSe was observed (highlighted in red) with complete disappearance of the γ-GeSe signal (**Figure 1g**). To compare the crystal orientation before and after the phase transition, the green dashed lines from **Figure 1f** were copied into **Figure 1g**, which reveals that the ZZ direction of α-GeSe aligns with one of the ZZ directions of the original γ-GeSe. The ZZ lattice constants of γ-GeSe and α-GeSe are 3.73 Å and 3.83 Å, respectively, indicating a small lattice mismatch along this direction. In contrast, due to the highly anisotropic nature of α-GeSe, other crystallographic directions exhibit significantly larger mismatches. Additionally, the in-plane structure of α-GeSe can be regarded as a distorted hexagonal structure, which can be derived by distorting γ-GeSe along the AC direction. This combination of minimal mismatch along the ZZ direction and structural compatibility likely facilitates a phase transition pathway that preserves alignment along the ZZ direction.

We also demonstrated the localized phase patterning from γ-GeSe to α-GeSe by laser-induced heating. A 2.33 eV laser with approximately 10 mW power was used as a heating source to precisely control the transition region within the sample. The laser, with a 1 μm diameter spot, was irradiated at 0.25 μm intervals to create a lateral junction between γ-GeSe and α-GeSe. **Figure 2a** presents an optical image of the encapsulated GeSe sample after the laser-induced phase transition. Stripe patterns of transformed α-GeSe, indicated by red arrows, are visible, while the other areas remain as γ-GeSe. A Raman intensity map obtained from this sample demonstrates that the irradiated regions were successfully transformed into α-GeSe (**Figure 2b**). Additionally, the inset in **Figure 2a** shows the SAED pattern of the phase-patterned sample, confirming the alignment of the crystal orientation of γ-GeSe and α-GeSe.

The observed crystal orientation and crystallinity of the transformed α-GeSe depend on the laser irradiation conditions. **Figure S3** presents the TEM analysis of α-GeSe formed by laser irradiation at 0.5 μm intervals. The transformed α-GeSe is polycrystalline, consisting of multiple domains that are rotated 30 ° relative to each other (**Figure S3a**). **Figure S3b** is a false-color TEM image indicating each domain, obtained using TEM dark-field imaging. **Figures S3c** and **S3d** show TEM dark-field images obtained using the red-highlighted and the blue-highlighted diffraction peaks in **Figure S3a**, respectively. With laser irradiation at narrow intervals, the preformed α-GeSe serves as a nucleation site, resulting in the templated growth of α-GeSe with improved structural quality. In contrast, when the laser is irradiated at wider intervals, independent nucleation of α-GeSe from γ-GeSe may occur, which can lead to polycrystalline α-GeSe.

Heterostructures between different materials or phases are important because they can exhibit new properties arising from their interfaces. In group-IV chalcogenides, previous studies have primarily focused on creating heterostructures through stacking, phase separation, or multi-step synthesis.[34,40-47] The utilized local phase patterning is an excellent way to form heterointerface between γ-GeSe and α-GeSe with well-aligned crystal orientation. For various TEM measurements, we transferred the GeSe sample with stripe-patterned α-GeSe onto a TEM grid (**Figure 2c**). To examine the atomic structure of this sample, we acquired atomic-resolution scanning transmission electron microscopy (STEM) images. γ-GeSe and α-GeSe maintained their well-known crystal structures away from the interface, with their crystal orientations aligned with each other (**Figure 2d, e**). **Figure 2f** shows the STEM image taken at the interface between γ-GeSe and α-GeSe. Due to the sample thickness (~60 nm), imaging the interfacial structure between γ-GeSe and α-GeSe at atomic resolution was challenging. We found that the focal heights for γ-GeSe and α-GeSe at the interface were different (**Figure S4a, b**). Additionally, fast Fourier transform (FFT)-filtered images of γ-GeSe and α-GeSe confirmed that the signal from γ-GeSe and α-GeSe overlap at the interface as shown in **Figure S4c** and **Figure S4d**. These results indicate that the junction between γ-GeSe and α-GeSe is not vertically straight (**Figure S4e**). The non-uniform heat generation and dissipation along the vertical direction during laser irradiation are likely responsible for the observed junction formation.

We performed electron energy loss spectroscopy (EELS) at the interface to analyze the electronic structure of the sample. **Figure 2g** displays the γ-GeSe/α-GeSe interface, where we obtained an EELS map from the region indicated by the yellow box. The main peak in the low-loss region corresponds to a plasma oscillation of the valence electrons in the material. In general, the measured plasmon energy $E_p$ can be described by the free-electron approximation, given by:

$$E_p = \hbar\sqrt{\frac{ne^2}{m\varepsilon_0}}, \quad (1)$$

where $\hbar$ is the reduced Planck constant, $n$ is the valence electron density, $e$ is the elementary charge, $m$ is the electron mass, and $\varepsilon_0$ is the vacuum permittivity.[48] Assuming that 10 valence electrons per formula unit contribute to the plasma oscillation, the calculated plasmon energies are 17.06 eV for γ-GeSe and 17.38 eV for α-GeSe. Experimentally, the plasmon energies are measured to be 17.85 eV for γ-GeSe and 17.80 eV for α-GeSe, consistently showing a slightly higher value for γ-GeSe across the mapping region **(Figure 2h, i)**. While the overall values are in good agreement with the free-electron model, the relative trend between the two phases is reversed: the calculation predicts a higher plasmon energy for α-GeSe, whereas the experiment shows a marginally higher energy for γ-GeSe. This inversion suggests that factors beyond the free-electron approximation—such as differences in band structures arising from distinct crystal structures—may influence the plasmon response in these materials. Further evidence of differing electronic structures can be observed in the weak energy-loss features shown in the inset of **Figure 2i**. γ-GeSe exhibits a relatively sharp peak near 7 eV, while α-GeSe displays a broad spectral feature spanning 5.5 to 11 eV. The sharp peak in γ-GeSe may originate from a gap between split sub-bands, which suppresses interband transitions at that energy.[49] Density of states (DOS) calculations support this interpretation, showing sharp and split features in γ-GeSe compared to the broader DOS observed in α-GeSe, consistent with the EELS measurements.[32]

We also measured the electrical properties of the γ-GeSe/α-GeSe heterostructure created via laser-induced phase transition. For device measurements, we used h-BN as an encapsulation layer and fabricated the devices using e-beam lithography (**Figure S5a**). After fabricating the h-BN-encapsulated device, we irradiated a laser onto a localized channel region, inducing the phase transition from γ-GeSe to α-GeSe (**Figure S5b**). **Figure S5c** presents the transfer curve of the γ-GeSe/α-GeSe heterostructure device, where the expected *p*-type characteristics of α-GeSe were observed. The transformation of the γ-GeSe device channel into α-GeSe facilitates the formation of heterojunctions between semiconducting and highly conducting materials.

To elucidate the underlying mechanism of the phase transition from γ-GeSe to α-GeSe, we gradually increased the annealing temperature and observed changes in the encapsulated sample. **Figure 3a** shows the encapsulated γ-GeSe before thermal annealing. After annealing at 540 °C, dark regions appeared near the edge of the sample (**Figure 3b**). A Raman spectrum confirmed that the central region remained as γ-GeSe (**Figure 3c**). On the other hand, a Raman spectrum obtained from the dark region near the edge revealed an additional peak at 210 $cm^{-1}$, characteristic of $GeSe_2$.[50-52] To confirm the formation of the $GeSe_2$ phase, we performed SAED acquisition and STEM imaging. The SAED exhibited not only the hexagonal symmetry of γ-GeSe but also peaks corresponding to $GeSe_2$, confirming its presence in the sample (**Figure 3d**). Furthermore, the crystal orientation of $GeSe_2$ was well-aligned with that of γ-GeSe. **Figure 3e** presents a STEM image of γ-GeSe with $GeSe_2$ formed at the edge of the sample, where contrast variations distinguish between the two regions. The green lines in the figure represent the ZZ-terminated edges of γ-GeSe. The predominance of ZZ-terminated edges suggests that $GeSe_2$ formation preferentially occurs in a specific direction of γ-GeSe. Additionally, atomic-resolution STEM imaging revealed an atomically well-defined interface between γ-GeSe and $GeSe_2$, possibly with covalent interfacial bonding (**Figure 3f**).

We further increased the annealing temperature until the phase transition was completed. After annealing at 555 °C, the encapsulated sample was fully transformed into α-GeSe, expect for certain regions exhibiting distinct contrast in the optical image (**Figure 3g**). As shown in the inset of **Figure 3g**, the dark regions in **Figure 3b** slightly expanded, and additional dark areas emerged. A Raman spectrum obtained from the central area displayed the characteristic spectrum of α-GeSe, while a Raman spectrum obtained from a dark region showed peaks of both α-GeSe and $GeSe_2$ (**Figure 3h**). A Raman intensity map of the sample further illustrates the spatial distribution of α-GeSe and $GeSe_2$ (**Figure 3i**). These results indicate that the sample underwent a complete transformation from γ-GeSe to α-GeSe, with residual $GeSe_2$ regions corresponding to the dark areas observed in the optical image.

The phase transition to α-GeSe and formation of minor $GeSe_2$ can be attributed to Ge vacancy clustering within the sample. The synthesized γ-GeSe naturally contains Ge vacancies of approximately 4~5%, which help stabilize the crystal structure.[36,38] As a result, γ-GeSe exhibits a slightly Se-rich chemical composition. **Figure 3j** displays a Ge-Se phase diagram, where the green vertical line indicates the chemical composition of γ-GeSe.[53,54] At room temperature, the most stable state for this composition consists of α-GeSe and $GeSe_2$. Furthermore, the formation energy of α-GeSe is lower than that of γ-GeSe, indicating that α-GeSe is the more stable phase (**Figure 3k**).[32] Considering these points, annealing encapsulated γ-GeSe cause Ge vacancies to migrate and cluster within the sample, resulting in the formation of Se-rich $GeSe_2$.[55,56] As the Ge vacancies in γ-GeSe decreases, its structural stability diminishes, causing it to undergo a phase transition into the more stable α-GeSe.

We compared the electronic and optoelectronic properties of γ-GeSe and transformed α-GeSe. **Figure S6** illustrates the phase transition and device fabrication process for the GeSe samples. To ensure the current path is confined within the sample, we used h-BN as the bottom encapsulation layer, while graphite was chosen as the top encapsulation layer (**Figure S6a**). **Figure S6b** shows the optical image after the phase transition by global annealing, where we confirmed the complete transformation of γ-GeSe into α-GeSe by measuring the Raman spectrum of the sample (**Figure S6c**). Following the phase transition, we selectively removed the top graphite layer through reactive ion etching (RIE) using $O_2$ plasma and fabricated electronic devices using e-beam lithography (**Figure S6d, e**). Atomic force microscopy (AFM) measurements revealed a flat surface of α-GeSe with a thickness of approximately 74 nm (**Figure S6f**).

**Figure 4a** compares the electrical properties of γ-GeSe and α-GeSe. The γ-GeSe device exhibited no gate dependence due to the high level of *p*-doping (~$10^{21}$ $cm^{-3}$) in the sample, which is consistent with previous reports.[35,36] In contrast, the transport properties of α-GeSe displayed a sharp difference compared to γ-GeSe. The α-GeSe device showed a *p*-type gate response, where the current increased with negative back gate bias (**Figures 4a and 4b**). Devices fabricated from transformed α-GeSe typically exhibited a field-effect mobility in the range of 0.1 to 1 $cm^2/V \cdot s$ and an on/off ratio ($I_{on}/I_{off}$) of approximately $10^2$, consistent with previous reports.[33,57-59] We note that improved electrical performance of α-GeSe has been demonstrated, indicating that there is room for improvement in the quality of the α-GeSe crystal and its interface.[60,61] Notably, the observed electrical resistance of γ-GeSe and α-GeSe differs by 7 orders of magnitude. In addition to the significant difference in electrical resistance, γ-GeSe and α-GeSe also exhibit distinct photoresponse characteristics. Similar to its gate response, γ-GeSe showed no detectable photoresponse, whereas α-GeSe exhibited robust photoconductance under light illumination, consistent with previous reports (**Figure 4c, Figure S7a**).[33,34,62] The photocurrent in α-GeSe device increased with light intensity, and its response time and responsivity were comparable to those of other 2D materials (**Figure S7b-f**).[63,64] **Figure 4d** and **Table 1** summarizes the resistance ratio between the low-resistance state (LRS) and high-resistance state (HRS) of various materials exhibiting crystalline-to-crystalline phase transitions.[15,23,24,65-69] The high resistance contrast between GeSe polymorphs underscores their potential as phase-change materials for memory applications.

## Conclusion

We demonstrated the phase transition from γ-GeSe to α-GeSe through global annealing or local laser-induced heating. Thermal annealing resulted in the complete transformation of γ-GeSe into the single-crystalline α-GeSe while maintaining a well-aligned crystal orientation with the original γ-GeSe. A well-aligned γ-GeSe/α-GeSe interface was achieved via laser irradiation, and STEM EELS analysis revealed distinct physical properties between γ-GeSe and α-GeSe. We propose that the phase transition mechanism involves Ge vacancy clustering, the subsequent formation of $GeSe_2$, and the stabilization of the crystal structure, leading to the transformation from γ-GeSe to α-GeSe. Notably, electrical resistance contrast between γ-GeSe and α-GeSe was found to be on the order of $10^7$. Given their high crystallinity and significant resistance contrast, GeSe polymorphs show potential as candidates for phase-change applications utilizing a crystalline-to-crystalline phase transition.

## Methods

**γ-GeSe synthesis.** γ-GeSe crystals were grown by the chemical vapor deposition method using Au nanoparticles as catalysts.[35,70] The $SiO_2$/Si substrate was coated with a 2.5 nm thick layer of Au by thermal evaporation. ~5.0 mg of GeSe powder (99.99 %; ALB Materials) and ~30 mg of $I_2$ crystal (99.8 %; Samchun Chemicals) were placed in a quartz boat, and the growth substrate was placed downstream. We found that the addition of iodine suppresses the formation of the α-GeSe phase and promotes the synthesis of the γ-GeSe phase. After a quartz tube was evacuated with a rotary pump, argon gas (300 sccm) was flowed for 10 min to remove oxygen. While the argon flow was maintained, the furnace was heated at 100 °C for 15 min. The rotary pump was turned off, and the pressure of the furnace was maintained at 1 atm by a leak valve. The furnace was heated to 550 °C with Ar gas of 150 sccm (usual heating rate: 15 °C/min) and maintained for 30 min. The target growth substrate was in the temperature zone of 370 °C during the growth process. After the synthesis, the vacuum pump was turned on, and the furnace was cooled to room temperature.

**Sample preparation and characterizations.** Optical microscopy images were acquired using a Leica DM-750M microscope under visible light. AFM topography images and data were obtained using a Park Systems XE-7 under ambient conditions. Raman spectra were measured using a 532 nm laser with a Nanobase XPER RF under ambient conditions. For TEM analysis, γ-GeSe and α-GeSe were transferred onto a holey $Si_3N_4$ TEM grid by a dry transfer method using Polydimethylsiloxane (PDMS) support. STEM images, SAED patterns and EELS spectra were obtained using JEOL-2100PLUS or double Cs-aberration-corrected JEOL ARM-200F, operated at 200 kV.

**Phase transition.** Graphite and h-BN were mechanically exfoliated onto a 300 nm $SiO_2$/Si substrate using PDMS to encapsulate γ-GeSe. Encapsulated γ-GeSe was annealed in a vacuum furnace at $10^{-4}$ torr. The temperature was ramped up to the target value in 30 minutes maintained at target temperature and then allowed to cool naturally to room temperature. For the laser-induced phase transition, a 2.33 eV laser with 10 mW power was irradiated for 0.2 s via scanning mode under ambient condition.

**Device fabrication and measurements.** γ-GeSe was transferred to 300 nm $SiO_2$/Si substrate by dry transfer method using PDMS. Electrode patterning was performed using the standard e-beam lithography method (PIONEER Two, Raith GmbH), and the electrode metals (Pt 100nm) were deposited by DC sputtering.[71] Electrical measurements were performed using a parameter analyzer (Keithley 4200A-SCS) at pressures between $10^{-6}$ and $10^{-7}$ torr. The photoresponse of the devices was measured using light-emitting diodes with energies of 1.91, 2.33, and 2.94 eV.

**Notes**

The authors declare no competing financial interest.

**Acknowledgments**

This work was supported by the National Research Foundation of Korea(NRF) grant funded by the Korea government(MSIT) (RS-2025-00560649), Yonsei Signature Research Cluster Program of 2024 (2024-22-0004), Global - Learning & Academic research institution for Master's·PhD students, and Postdocs (G-LAMP) Program of the National Research Foundation of Korea grant funded by the Ministry of Education (No. RS-2024-00442483), and Samsung Electronics Co., Ltd (IO210202-08367-01).

**Supporting Information**

The Supporting Information is available free of charge at https://pubs.acs.org/doi/.

Details of the annealing conditions for the phase transition from γ-GeSe to α-GeSe, additional experimental data on the crystallinity of α-GeSe, the interface structure and transport properties of the γ-GeSe/α-GeSe junction, the device fabrication process, and the optoelectronic properties of α-GeSe.

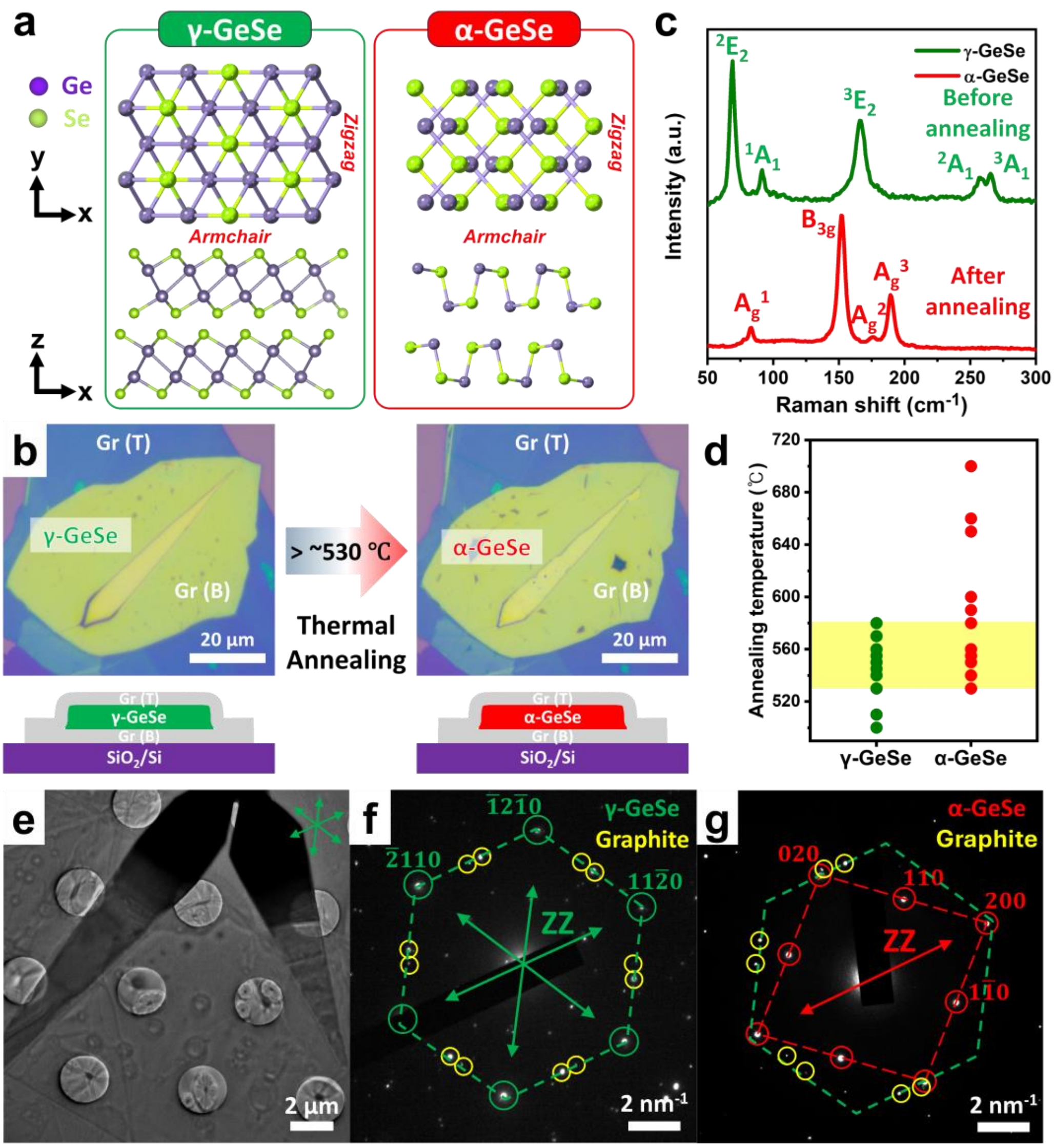

**Figure 1. Phase transition from γ-GeSe to α-GeSe via global annealing.** (a) Schematic representations of the crystal structures of γ-GeSe and α-GeSe, with plan-view (top) and side-view (bottom) illustrations. (b) Optical images (top) and corresponding schematics (bottom) depicting the phase transition from γ-GeSe to α-GeSe via encapsulation annealing. (c) Raman spectra acquired before and after the phase transition. (d) Temperature-dependent results of encapsulated samples after the thermal annealing. (e) TEM image of encapsulated γ-GeSe prior to the phase transition. Green arrows indicate the ZZ lattice directions. (f) SAED pattern of γ-GeSe before the phase transition, with green arrows marking the ZZ directions. (g) SAED pattern of α-GeSe after the phase transition, with red circles highlighting the diffraction spots and the red arrow exhibits the ZZ direction of the α-GeSe sample. For comparison, the SAED pattern of γ-GeSe before the phase transition is also shown.

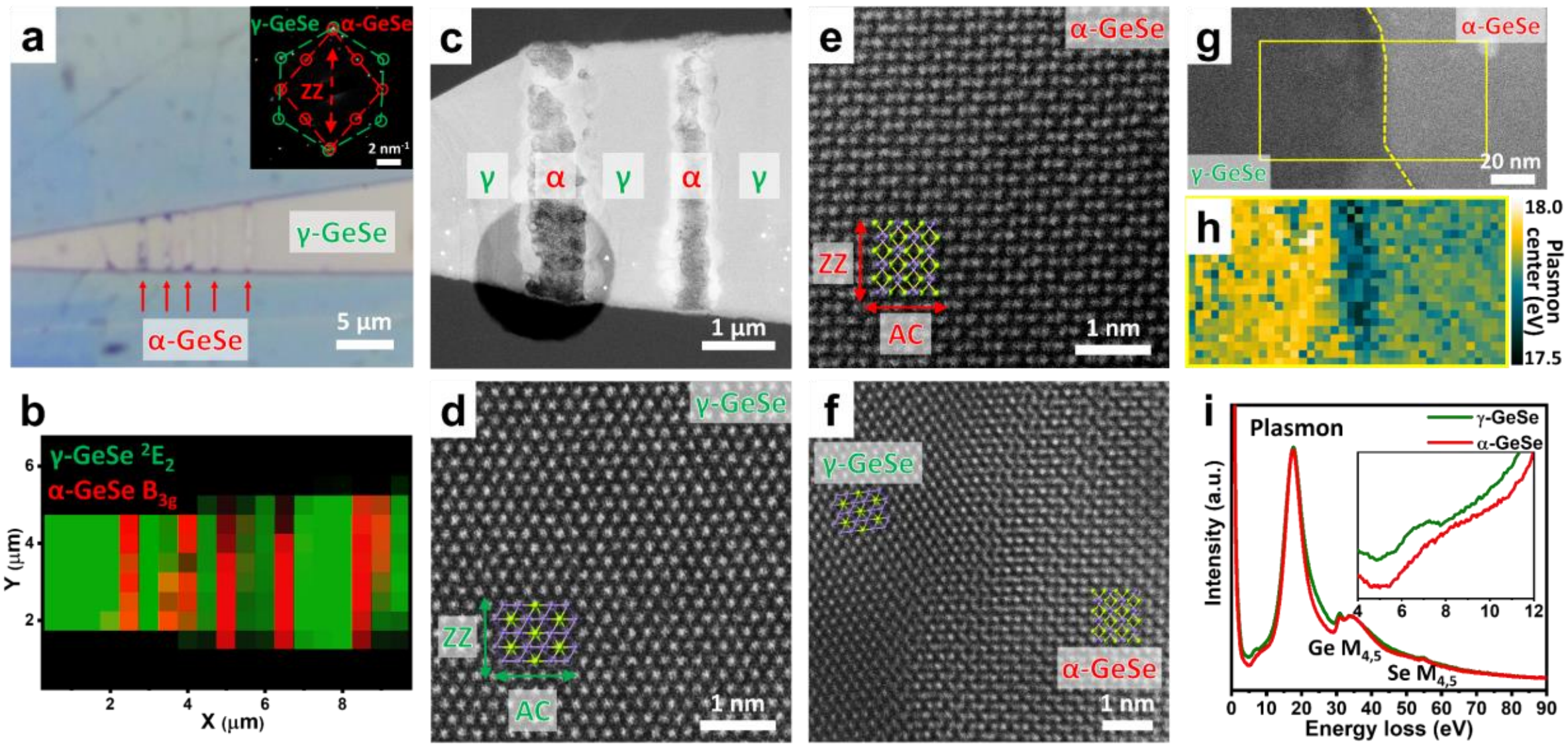

**Figure 2. Laser-induced patterned formation of α-GeSe from γ-GeSe.** (a) Optical image of the phase-patterned GeSe sample. The inset displays the SAED pattern of the sample, confirming a well-aligned crystal orientation. (b) Raman intensity map of the phase-patterned sample. (c) Low-magnification STEM image of the phase-patterned sample. (d) Atomic-resolution STEM image of γ-GeSe with an overlaid plan-view atomic structure. (e) Atomic-resolution STEM image of α-GeSe with an overlaid plan-view atomic structure. (f) Atomic-resolution STEM image of the γ-GeSe/α-GeSe interface. (g) STEM image near the interface. The yellow box indicates the field of view for EELS signal mapping in panel h. (h) Two-dimension mapping of the center position of the plasmon signal from the EELS measurement. (i) Exemplary EELS spectrum of γ-GeSe and α-GeSe, with the inset highlighting the weak energy loss structure.

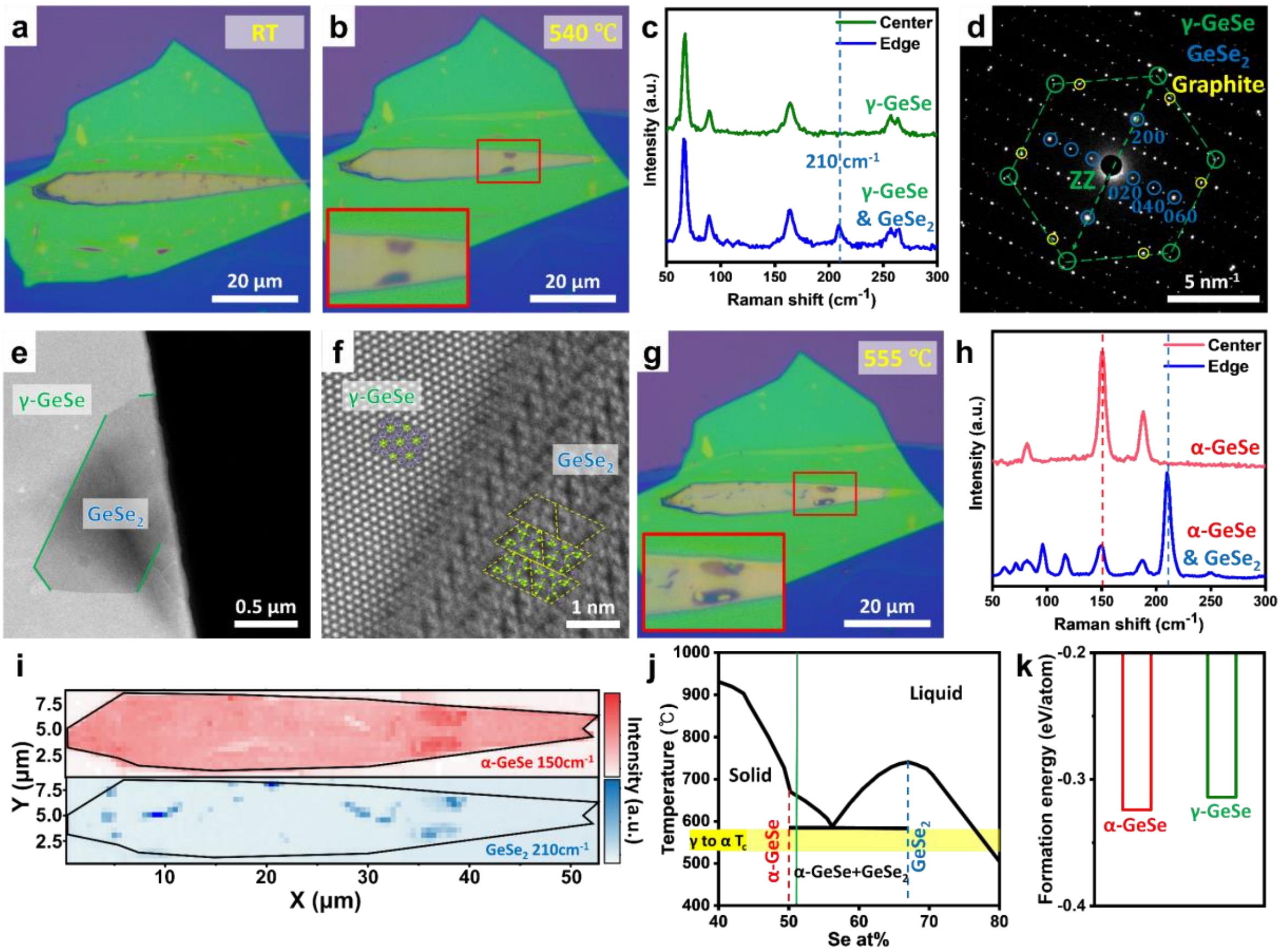

**Figure 3. Phase transition mechanism.** (a) Optical image of encapsulated γ-GeSe before annealing. (b) Optical image of γ-GeSe after annealing at 540 °C, with the inset highlighting dark regions formed after annealing. (c) Raman spectra from the center and the dark region near the sample edge. (d) SAED pattern of the sample near the edge after the annealing, where the green and blue diffraction signals correspond to γ-GeSe and $GeSe_2$ phases, respectively. The green arrow indicates the ZZ direction of γ-GeSe. (e) STEM image of the sample near the edge after annealing at 540 °C, showing both γ-GeSe and $GeSe_2$. Green lines indicate ZZ-terminated edges of γ-GeSe. (f) Atomic-resolution STEM image of the γ-GeSe/$GeSe_2$ interface. (g) Optical image of the sample after annealing at 555 °C, with the inset showing the expansion of dark regions. (h) Raman spectra from the center and edge regions after the phase transition to α-GeSe. (i) Raman intensity map of the sample, based on the peaks at 150 and 210 $cm^{-1}$. (j) Ge-Se phase diagram with the green line indicating the expected chemical composition of γ-GeSe. (k) Comparison of formation energies between α-GeSe and γ-GeSe.

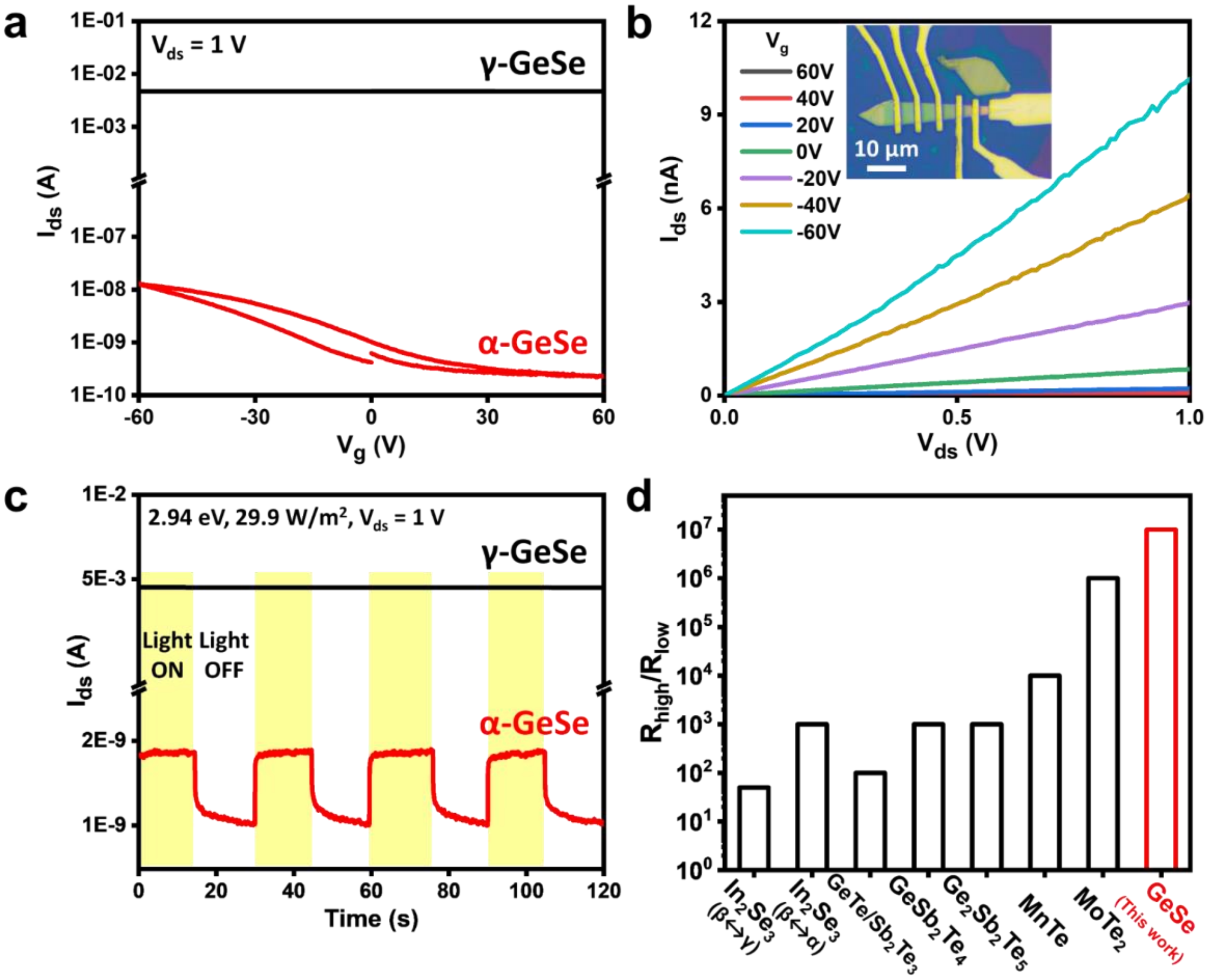

**Figure 4. Electrical and optoelectronic properties of GeSe devices before and after the phase transition.** (a) Transfer curves of γ-GeSe and α-GeSe devices, illustrating the significant changes in electrical conductivity after the transition. (b) Representative gate-dependent output curves of the fully-converted α-GeSe device by global annealing. (c) Photoresponse characteristics of γ-GeSe and α-GeSe devices. (d) Comparison of resistance contrasts observed in various crystalline-to-crystalline phase transitions.

| LRS | HRS | $R_{high}/R_{low}$ | Reference |
|---|---|---|---|
| β-$In_2Se_3$ | γ-$In_2Se_3$ | 50 | [65] |
| β-$In_2Se_3$ | α-$In_2Se_3$ | ~$10^3$ | [66] |
| GeTe/$Sb_2Te_3$ - SET | GeTe/$Sb_2Te_3$ - RESET | ~$10^2$ | [24] |
| h-$GeSb_2Te_4$ | c-$GeSb_2Te_4$ | ~$10^3$ | [67] |
| h-$Ge_2Sb_2Te_5$ | c-$Ge_2Sb_2Te_5$ | ~$10^3$ | [68] |
| α-MnTe | β-MnTe | ~$10^4$ | [69] |
| 1T'-$MoTe_2$ | 2H-$MoTe_2$ | ~$10^6$ | [15] |
| **γ-GeSe** | **α-GeSe** | **~$10^7$** | **This work** |

**Table 1. Summary of resistance contrast of crystalline-to-crystalline phase transitions among chalcogenides.**